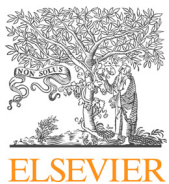

Contents lists available at ScienceDirect

# Science Bulletin

journal homepage: www.elsevier.com/locate/scib

Article

# Demonstration of topological wireless power transfer

Li Zhang [a,b,c,1], Yihao Yang [a,b,c,d,e,\*,1], Zhao Jiang [f], Qiaolu Chen [a,b,c], Qinghui Yan [a,b,c], Zhouyi Wu [f], Baile Zhang [d,e,\*], Jiangtao Huangfu [f,\*], Hongsheng Chen [a,b,c,\*]

[a] *Interdisciplinary Center for Quantum Information, State Key Laboratory of Modern Optical Instrumentation, College of Information Science and Electronic Engineering, Zhejiang University, Hangzhou 310027, China*
[b] *ZJU-Hangzhou Global Science and Technology Innovation Center, Key Laboratory of Advanced Micro/Nano Electronic Devices & Smart Systems of Zhejiang, Zhejiang University, Hangzhou 310027, China*
[c] *International Joint Innovation Center, ZJU-UIUC Institute, Zhejiang University, Haining 314400, China*
[d] *Division of Physics and Applied Physics, School of Physical and Mathematical Sciences, Nanyang Technological University, Singapore 637371, Singapore*
[e] *Centre for Disruptive Photonic Technologies, The Photonics Institute, Nanyang Technological University, Singapore 639798, Singapore*
[f] *Laboratory of Applied Research on Electromagnetics (ARE), Zhejiang University, Hangzhou 310027, China*

**ABSTRACT**

Recent advances in non-radiative wireless power transfer (WPT) technique essentially relying on magnetic resonance and near-field coupling have successfully enabled a wide range of applications. However, WPT systems based on double resonators are severely limited to short- or mid-range distance, due to the deteriorating efficiency and power with long transfer distance. WPT systems based on multi-relay resonators can overcome this problem, which, however, suffer from sensitivity to perturbations and fabrication imperfections. Here, we experimentally demonstrate a concept of topological wireless power transfer (TWPT), where energy is transferred efficiently via the near-field coupling between two topological edge states localized at the ends of a one-dimensional radiowave topological insulator. Such a TWPT system can be modelled as a parity-time-symmetric Su-Schrieffer-Heeger (SSH) chain with complex boundary potentials. Besides, the coil configurations are judiciously designed, which significantly suppress the unwanted cross-couplings between nonadjacent coils that could break the chiral symmetry of the SSH chain. By tuning the inter- and intra-cell coupling strengths, we theoretically and experimentally demonstrate high energy transfer efficiency near the exceptional point of the topological edge states, even in the presence of disorder. The combination of topological metamaterials, non-Hermitian physics, and WPT techniques could promise a variety of robust, efficient WPT applications over long distances in electronics, transportation, and industry.

## 1. Introduction

The last decade has witnessed a rapid development of non-radiative wireless power transfer (WPT) using a pair of magnetically coupled resonators in the near field (Fig. 1a), which could date back to Nikola Tesla's pioneering work over a century ago [1–5]. Such a two-resonator WPT system can achieve efficient energy transfer in short- or mid-range distance (i.e., the transmission distance is less than or comparable to the resonator's dimension), and has successfully found a variety of applications ranging from electric vehicles [6] to biomedical implants [7]. However, the coupling strength between the transmitter and receiver decreases dramatically as the transmission distance increases, leading to a significant drop in both the power and efficiency.

In order to extend the transfer distance without sacrificing the efficiency, multi-relay coils have been added between the source and receiver resonators, forming a domino-form coil chain [8–10]. This coil-resonator chain is a magneto-inductive waveguide where the energy is transferred via guided modes (Fig. 1b). Unfortunately, the previous domino-form WPT systems suffer from several limitations. On one hand, due to the complex cross-coupling between nonadjacent coils, the optimal operating frequency shifts away from a single coil's resonance frequency and depends sensitively on the numbers of the coils in the resonator chain [8]. On the other hand, there exist numerous standing-wave modes in a coil-resonator chain, which are close to each other in frequency. Therefore, the mode used to transfer energy is very likely to couple with other modes due to a small perturbation (such as changes of the

⇑ Corresponding authors.
*E-mail addresses:* yangyihao@zju.edu.cn (Y. Yang), blzhang@ntu.edu.sg (B. Zhang), huangfujt@zju.edu.cn (J. Huangfu), hansomchen@zju.edu.cn (H. Chen).
[1] These authors contributed equally to this work.

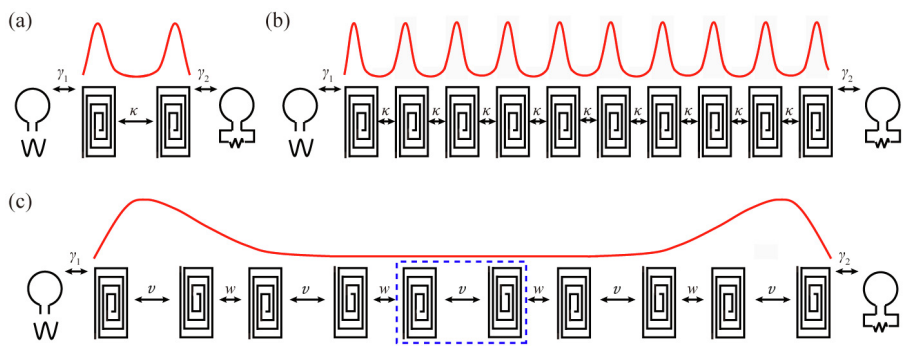

**Fig. 1.** (Color online) Comparison between WPT with two-coils, WPT with domino-form coils, and TWPT. (a) WPT with two coils. A wave is emitted by the generator and then coupled to the coils with a rate $\gamma_1$. By magnetic coupling coefficient $\kappa$, the energy is transmitted to the left coil and then coupled to the load with a rate $\gamma_2$. The transferred energy is localized on the two coils (red line). (b) Domino-form WPT scheme with multi-relay coils placed with equal distances. The power is emitted from the generator and then coupled to the first coil with a rate $\gamma_1$. By magnetic coupling coefficient $\kappa$, the energy propagates through the multi-relay coils, and then transfers to a load with a rate $\gamma_2$. The energy is transferred via standing-wave modes, superposition of two counter-propagating modes (red line). (c) TWPT scheme in this work. The power is generated at the source and transfers to the first coil with a rate $\gamma_1$. Via the near-field coupling with coefficients $v$ and $w$, the energy transfers to multi-relay coils and then taken out at the load with a rate $\gamma_2$. The energy is localized on the two ends and decays into the middle (red line). Each unit cell (blue box) is composed of two coils with opposite winding directions.

distances between coils), leading to a dramatic drop and fluctuation of the transmission power and efficiency. Realization of a WPT device that can transfer energy efficiently over a relatively long distance and robustly deliver power against perturbations caused by the unavoidable fabrication imperfections is highly desired, which, however, remains elusive.

Recently, tremendous efforts have been devoted to topological photonics, in which band topology theory is introduced into photonics [11–24]. The most striking feature of topological photonic materials is that though their bulks are opaque, their edges support topological boundary states that are robust against defects and disorders [25,26]. The topological edge states have enabled many promising applications, such as robust optical delay lines [14], topological lasers [27,28], and robust on-chip communications [29]. Inspired by these intriguing properties, it is natural to ask whether the topology can be applied to WPT to design a topologically robust WPT device whose performance is immune to perturbations and fabrication imperfections.

Here, we experimentally demonstrate the concept of topological wireless power transfer (TWPT), where the energy is transferred efficiently via the near-field coupling between two topological edge states localized at the ends of a one-dimensional (1D) radiowave topological insulator. The structure consists of a coil-resonator chain (Fig. 1c), where two coil resonators comprise a unit cell (blue dashed box), and the intra- and inter-cell couplings are tunable. This is an exact Su-Schrieffer-Heeger (SSH) model [30] (without considering the source and load). Besides, the winding directions of two adjacent coils in each unit cell are opposite, which strongly suppresses the unwanted cross-couplings between nonadjacent coils that could break the chiral symmetry of the SSH model. Based on the coupled-mode theory (CMT) [31], the source and load can be viewed as two conjugated imaginary potentials at the end of the coil-resonator chain, and, therefore, the whole TWPT system can be modelled as a parity-time-symmetric (PT-symmetric) non-Hermitian SSH chain with complex boundary potentials. Furthermore, we theoretically and experimentally demonstrate that by tailoring the inter- and intra-cell coupling strengths, the whole TWPT system can work at the exceptional point of the topological edge states, with maximum energy efficiency around the resonance frequency of a single-coil resonator. Remarkably, due to the topological nature of the edge states, the TWPT system can maintain relatively high efficiency at a fixed operating frequency in the presence of disorder, in comparison to the conventional domino-form WPT system. Note that though it was proposed that the SSH chain [32] or other 1D topological models (such as Harper chain [33]) may find applications in WPT, experimental demonstration of the TWPT system has never been done before.

## 2. Methods

### 2.1. Experimental samples

The TWPT samples are fabricated with printed circuit board (PCB) technology by coating a 0.035-mm-thick copper film on a 2-mm-thick dielectric F4B substrate (relative permittivity 3). The parameters of the resonator are inner radius $r = 40$ mm, coil width $w = 6$ mm, gap $g = 3$ mm, and number of turns $N = 12$, as illustrated in Fig. S1a (online). Two plastic pillars (radius 6 mm) are inserted through air holes (radius 6 mm) at the corners of each board, to make all boards aligned. The coil resonator chain consists of ten resonators where two neighbor coils have opposite winding directions. In the measurement, each layer can slide coaxially along the $x$-direction, as shown in the Supplementary materials Section S1.

### 2.2. Numerical simulations

The band diagrams of the unit cells composed of coils with the same/opposite winding directions are numerically calculated in the eigenvalue module of the commercial software Computer Simulation Technology (CST) Microwave Studio. Here, for simplicity, the metallic coils are modelled as a perfectly electric conductor (PEC), and the loss tangent of the dielectric substrate is neglected.

### 2.3. Measurements

The experimental setup is shown in Fig. S2a (online). Two small copper coils connected to a vector network analyzer (VNA) (R&S ZNB 20) are employed to measure the efficiency spectrum. The experimental field distributions are obtained by measuring the local field of each coil resonator of the TWPT chain with a probe.

### 2.4. Powering a light-emitting diode with the TWPT system

As illustrated in Fig. S2b (online), the experiment setup includes an alternating current (AC) power generator, an antenna tuner, a light-emitting diode (LED) circuit board, two copper coil resonators (acting as the source and receiver resonators), and the TWPT chain. The AC power generator and the antenna tuner are connected to the source resonator with a matched impedance. The LED circuit board, composed of a rectifier circuit (converting AC to direct current (DC)), two capacitors, and a 1-W white LED, is connected with the receiver resonator (Fig. S3 online). The operating frequency of the power supply is tuned to 12.78 MHz.

## 3. Results

As depicted in Fig. 1c, the designed TWPT system consists of a source, a coil-resonator chain, and a receiver load. The coil-resonator chain, behaving as a finite SSH lattice model, is composed of $N$ coil resonators with the same resonance frequency $\omega_0$ and the same intrinsic loss $\Gamma_0$. Two adjacent coils are coupled to each other with magnetic coupling coefficients $v$ and $w$, for intra- and inter-cell hopping, respectively. The coupling coefficient between the source (receiver) and the side coil resonator is $\gamma_1 = \gamma$ ($\gamma_2 = \gamma$). Here, the cross-coupling between nonadjacent coils is not considered. According to the CMT, [31] the dynamics of this system is described by the following equations,

$$\begin{aligned} da_1/dt &= (-i\omega_0 - \Gamma_0 - \gamma)a_1 - iva_2 + \sqrt{2\gamma}s_{1+}, \\ da_2/dt &= (-i\omega_0 - \Gamma_0)a_2 - iva_1 - iwa_3, \\ da_N/dt &= (-i\omega_0 - \Gamma_0 - \gamma)a_N - iva_{N-1}. \end{aligned} \quad (1)$$

Here, $a_m = A_m e^{-i\omega t}$ denotes the resonance modes of the $m$th relay and $s_{1+}$ represents the incident wave. When the return wave $s_{1-}$ equals zero, we can obtain the eigenfrequencies of the system [34]. The dynamics in the system can be represented as $H\mathbf{V} = \omega \mathbf{V}$, where $\mathbf{V} = (a_1, a_2, \dots a_N)^T$. To simplify the system, the intrinsic loss $\Gamma_0$ is omitted, which acts as a lossy background that does not affect our conclusions [35,36] (see the Supplementary materials Section S2). Then we obtain an effective Hamiltonian $H$ of this non-Hermitian system,

$$H = \begin{bmatrix} \omega_0 + i\gamma & v & 0 & \dots & 0 & 0 \\ v & \omega_0 & w & \dots & 0 & 0 \\ 0 & w & \omega_0 & \dots & 0 & 0 \\ \dots & \dots & \dots & \dots & \dots & \dots \\ 0 & 0 & 0 & \dots & \omega_0 & v \\ 0 & 0 & 0 & \dots & v & \omega_0 - i\gamma \end{bmatrix}. \quad (2)$$

Here, $+\gamma$ ($-\gamma$) denotes the effective gain (loss) strength. This Hamiltonian is a finite non-Hermitian SSH lattice Hamiltonian with two conjugated imaginary potentials at both ends. Noted that, due to $[PT, H] = 0$ ($P$ and $T$ are the parity operator and the time-reversal operator, respectively), $H$ is PT-symmetric [30,37].

When $\gamma = 0$, $H$ is a Hermitian SSH Hamiltonian. The SSH chain can be viewed as a 1D topological insulator, whose topological property is characterized by the celebrated Zak phase [38] $\varphi_{Zak} = i \int_{-\pi/p}^{\pi/p} <\mu_{n,k}|\partial_k|\mu_{n,k}> dk$. Here $p$ is the period of the SSH chain and $\mu_{n,k}$ denotes the periodic in-cell part of the normalized Bloch eigenfunction of a state in the $n$th band with wavevector $k$. For $w/v > 1$, the Zak phase is $\pi$, indicating a topologically nontrivial phase (see the yellow region in Fig. 2a); for $w/v < 1$, the Zak phase is 0, leading to a topologically trivial phase [39] (see the white region in Fig. 2a).

Now we consider a finite TWPT chain with size $N = 10$, resonance frequency $\omega_0 = 12.78$ MHz, and effective gain/loss strength $\gamma = 0.5$ MHz. According to Eq. (2), one can obtain the real and imaginary part of the corresponding eigenfrequencies as a function of $w/v$, as shown in Fig. 2a and b, respectively. In the case of $w/v < 1$, a trivial bandgap opens around $\omega_0$. When $w/v > 1$, a topologically nontrivial bandgap opens around $\omega_0$, in which two topological edge states (the fifth and sixth modes in Fig. 2a) emerge. In an infinitely-long Hermitian SSH chain, the eigenfrequencies of the two topological edge states should degenerate exactly at $\omega_0$ for $w/v > 1$. However, in an SSH chain with finite length, the two topological edge states couple to each other via near-field coupling, rendering frequencies of the edge states deviate from $\omega_0$. Besides, due to the gain/loss effect contributed by the source/load, the whole system involving two coupled edge states can be viewed as a pair of PT-symmetric coupled resonators, resulting in a PT-symmetric phase transition [30,40]. At the critical point of $w/v = \kappa_c$, or the exceptional point, [41] both the eigenfrequencies and eigenmodes of two edge states coalesce. In the exact PT-symmetry phase ($w/v < \kappa_c$), the eigenfrequencies of two edge states are purely real but split away from $\omega_0$. In the broken PT-symmetry phase ($w/v > \kappa_c$), the eigenfrequencies of two edge states become complex while their real parts preserve $\omega_0$ (solid colored lines). The other eight eigenfrequencies of bulk states remain real during the phase transition (dashed colored lines).

We then calculate the efficiency of WPT over the SSH chain based on the CMT (see details in the Supplementary materials Section 3), as shown in Fig. 2c. In the exact PT-symmetry phase regime ($w/v < \kappa_c$), we can achieve the maximum efficiency around two edge-state eigenfrequencies that deviate away from center frequency $\omega_0$. In the broken PT-symmetry phase regime ($w/v > \kappa_c$), the efficiency reaches the maximum at $\omega_0$, while the maximum efficiency is relatively low and decreases rapidly as $w/v$ increases. At the exceptional point ($w/v = \kappa_c$), the efficiency can be as high as unity at $\omega_0$. Impressively, these three situations are similar to the over coupling, under coupling, and critical coupling cases in the well-known double-coil WPT system [3].

We argue that the TWPT system relying on the near-field coupling of the topological edge states is superior to the conventional domino-form WPT system based on magneto-inductive waveguiding modes, especially in the presence of disorder, due to the topologically robust nature of the edge states. To verify the above statement, we compare the calculated efficiencies of both nontopological and topological WPT systems in the presence of disorder. The evolution of the efficiency spectrum at a fixed frequency as a function of disorder strength is illustrated in Fig. 2d. Here, the disorder is introduced by randomly changing the coupling coefficients $v$ and $w$. Each case is averaged over 1000 realizations. Note that for the TWPT system ($w/v = \kappa_c$), the fixed operating frequency is chosen at $\omega_0$; for the conventional domino-form WPT system ($w/v = 1$) that is similar to the equidistance coil-resonator system, the operating frequency is selected at the middle peak of the eigenfrequency spectrum (see the Supplementary materials Section 4). Fig. 2d presents that in comparison to the nontopological WPT system, the TWPT system always maintains significantly higher efficiency at different disorder strengths. We should note that though the edge states are topologically robust against disorders, the coupling between these two states may change as the disorder strength varies. This can explain why the efficiency of the TWPT system does not maintain unity as the disorder strength increases.

Next, we perform experiments to corroborate the high-efficiency TWPT system. As shown in Fig. 3a, each coil has width $w = 6$ mm, gap $g = 3$ mm, and inner radius $r = 40$ mm, which is fabricated by printing 0.035-mm-thick copper cladding onto a 2-mm-thick F4B PCB with relative permittivity 3 (see the Supplementary materials Section 1). There are ten coils arranged on the $x$-direction which can slide coaxially at will. Each unit cell (see the black dashed box in Fig. 3a) is composed of two coils with a period being $p = 17$ cm. The distance between two coils in a unit cell is $p/2 + s$, where $s$ denotes the shift that determines the inter- and intra-cell coupling $w$ and $v$. Note that adjacent coils have opposite winding directions (indicated by the blue or red arrows in Fig. 3a), which is distinct from the conventional domino-form WPT system with all coils having the same winding direction. By doing so, the unwanted cross-couplings between nonadjacent coils are strongly suppressed (see the Supplementary materials Section 5). The unpleasant cross-couplings usually make the optimal operating frequency shift away from the resonance frequency of a single-coil resonator and also render the operating frequency depend sensitively on the numbers of the coils [8]. Besides, the cross-couplings usually break the chiral symmetry of the SSH chain that protects the edge states [39]. We also numerically calculate the band structures of our design and the coil chain with the same winding direction in a commercial software CST Microwave Studio, respectively (see Methods). As shown in Fig. 3b, the band structure of our design (black lines) is indeed symmetric with respect to the resonance frequency of a single resonator (around 13 MHz), indicating the preserved chiral symmetry [39]. In contrast, the band structure of the coil chain with the same winding direction (magenta lines) is asymmetric, implying the chiral symmetry breaking.

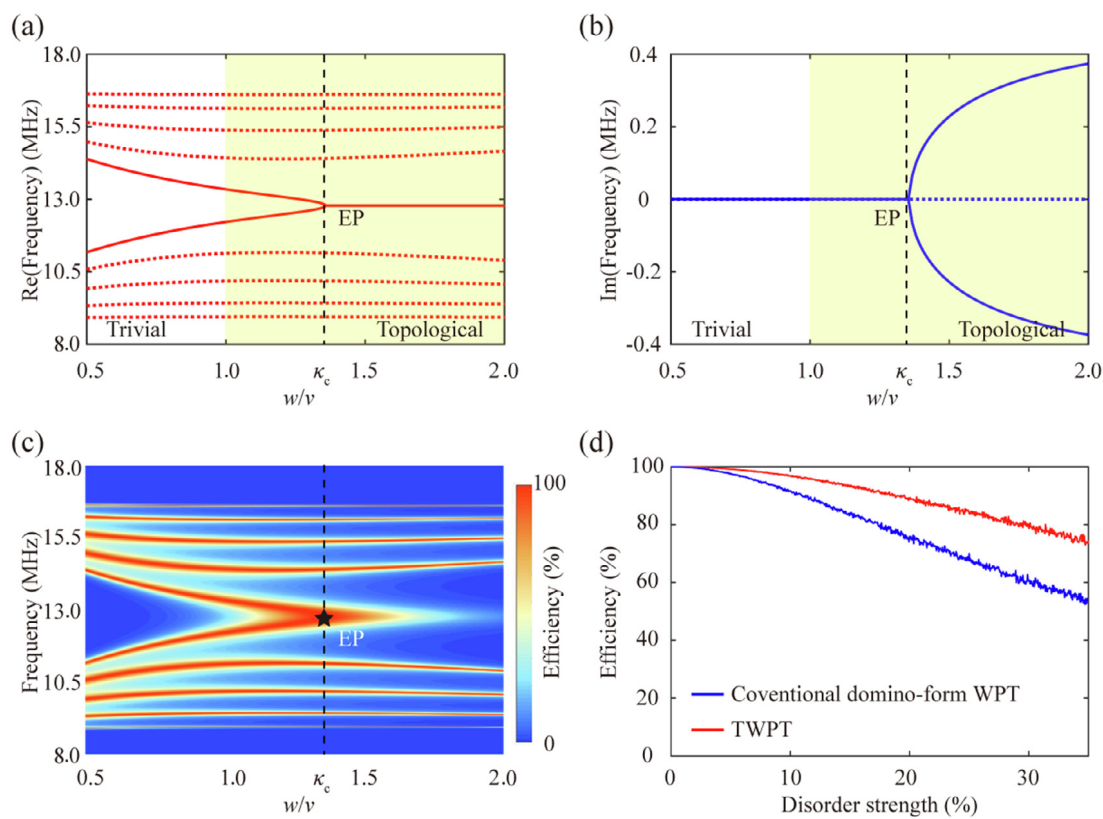

**Fig. 2.** (Color online) Theoretical analysis. (a, b) The real and imaginary part of the eigenfrequencies. The yellow and white regions indicate the topologically nontrivial phase and the trivial phase, respectively. The dashed black line denotes the exceptional point at $w/v = \kappa_c$. The solid and dashed colored lines indicate the edge modes and the bulk modes, respectively. (c) Numerically calculated efficiency spectrum versus frequency and $w/v$ based on CMT. (d) Averaged efficiency as a function of disorder strength in conventional domino-form WPT (blue line) and TWPT systems (red line).

To measure the efficiency spectrum of this TWPT system, we place two probe coils at the two sides of the coil chain connected with a VNA, serving as the source and the receiver, [4,5] respectively (see Methods). Fig. 3c and d represent the measured efficiency spectra as a function of frequency and $w/v$. This is consistent with the analytically-calculated efficiency spectrum shown in Fig. 2c. Note that the relation between $w/v$ and $s$ can be found in the Supplementary materials Section 6. When $w/v < 1$, the intra-cell coupling is larger than the inter-cell counterpart, hence no edge state is observed in the bulk bandgap, indicating a topologically trivial phase. By contrast, when $w/v > 1$, the intra-cell coupling is smaller than the inter-cell one, and thus two edge states exist in the bulk bandgap, implying a topologically nontrivial phase. Additionally, a PT phase transition occurs around the exceptional point of frequency $f_0 = 12.78$ MHz (magenta dashed line in Fig. 3c) and $w/v = 1.3$ ($s = 11$ mm), resulting from the gain-loss effect. One can see that around the exceptional point, the transmission efficiency reaches the maximum, to be specific, 60.2%. This is consistent with the analytically-calculated efficiency spectrum shown in Fig. 2c. Note that due to the metallic loss, the measured efficiency is not unity (see the Supplementary materials Section 7). We further compare the field distributions at different phases, as shown in Fig. 3e and f. In comparison to the broken-PT case, the mode energy in the exact-PT phase is distributed over the chain. Neither of the topological edge modes experiences net loss or gain. Note that the TWPT relies on an appropriate evanescent coupling between two topological edge states. Therefore, the topological edge states cannot be too localized, which may render the coupling to be too weak to transfer the energy.

Also, we use the TWPT system to power a LED load at the frequency of 12.78 MHz (Methods). The brightness of the LED measures the efficiency of the TWPT system directly. Interestingly, one can see that when increasing $s$ from 2 to 19 mm, the LED becomes brighter at first and then dimmer (see the Supplementary materials Section 8). The brightness of the LED reaches the maximum at precisely the exceptional point $s = 11$ mm. These phenomena are consistent with our analytical and measured efficiency spectrum.

We also experimentally compare the efficiency of the TWPT system with that of the conventional domino-form WPT system in the presence of disorder. Note that for the two systems without disorder (blue cycles and blue squares in Fig. 4), their maximum efficiencies are close, which is consistent with our theoretical calculation (see the Supplementary materials Section 4). The disorder is introduced by randomly moving the coil resonators away from their original positions by $\Delta s$, with $\Delta s \in [-d, d]$ being a uniformly distributed disorder. The resulting disordered configurations are shown in the top panel of Fig. 4, where the TWPT and nontopological WPT systems have the same disorder pattern. In Fig. 4a and b (Fig. 4c and d), *d* = 1.5 cm (2.5 cm), which is about 17% (30%) of the average distance between two neighbor coils. It is evident that the efficiency maintains high and the frequency shift of efficiency peak (<0.5%) is negligible in the TWPT system, owing to the topological protection. By contrast, the efficiency peak in the nontopological WPT system shifts significantly, and the efficiency at the operating frequency dramatically decreases. This is because the disorder causes detrimental coupling between the operating resonance with the other resonances. Additionally, comparing Fig. 4a with c (Fig. 4b and d), the corresponding maximum efficiency around the operating frequency (green dashed line) decreases as the disorder strength increases. These experimental observations are in good agreement with our theoretical analysis (Fig. 2d).

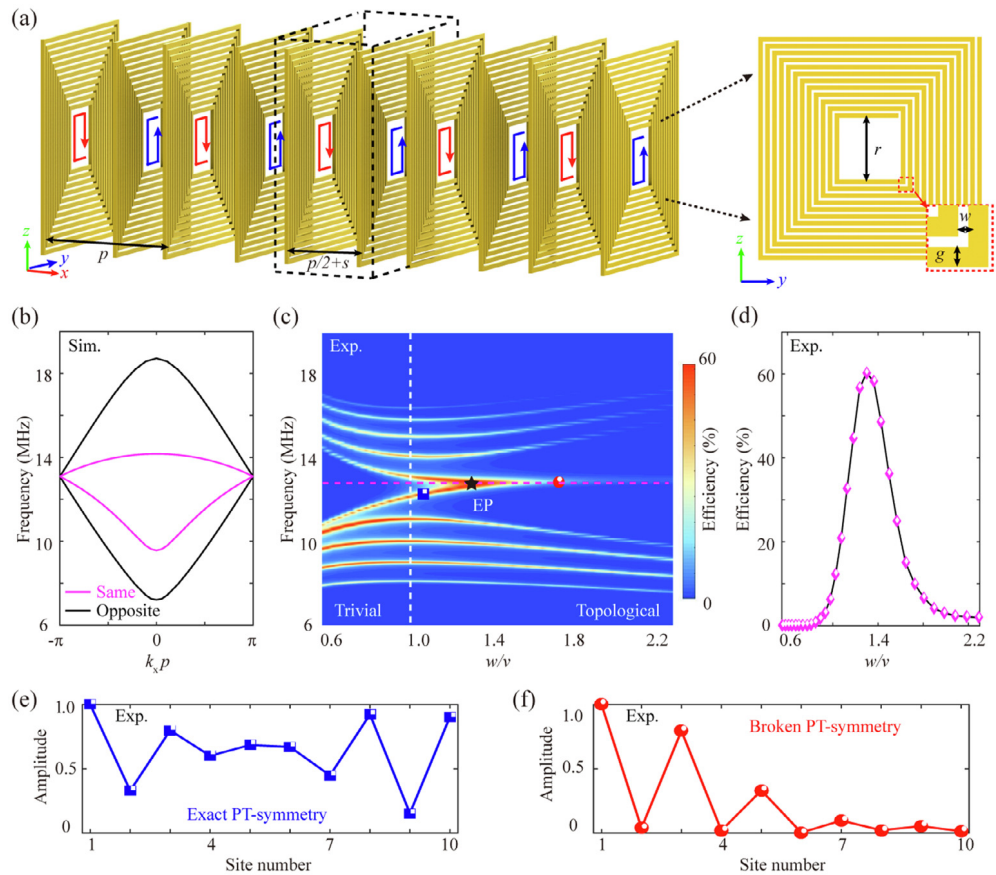

**Fig. 3.** (Color online) Experimental demonstration of the TWPT system. (a) Schematic of the TWPT chain. Ten coil resonators are placed along *x*-direction in a period *p* = 17 cm. Each unit cell (black dashed box) is composed of two coils with opposite winding directions (marked with red and blue arrows). The coils can slide coaxially at will and the distance between two coils in a unit cell is *p*/2 + *s*. For *s* = 0 mm, all the coils are placed with equal distances *p*/2. The right figure is the front view of one coil. The structure parameters are inner radius *r* = 40 mm, coil width *w* = 6 mm, gap *g* = 3 mm, and number of turns *N* = 12, respectively. (b) Band structure of unit cells composed of same/opposite winding directions with shift *s* = 0 mm and *p* = 17 cm, represented by magenta/black lines. (c) Measured efficiency versus both frequency and *w/v*. The white dashed line indicates *w/v* = 1, separating the topologically trivial and nontrivial phase. The magenta dashed line indicates the center frequency $f_0$ = 12.78 MHz. (d) Measured efficiency versus *w/v* at a fixed frequency $f_0$ = 12.78 MHz. (e, f) Experimentally achieved field distributions of this system in exact/broken PT phase (marked with blue squares/red cycles in (c)).

## 4. Conclusion

To conclude, we have thus experimentally demonstrated an efficient TWPT system based on the 1D radiowave topological insulator for long-distance transmission. The multiple relays in the TWPT system are judiciously designed to suppress the detrimental cross-coupling between nonadjacent coil resonators. The present TWPT system can be considered as a PT-symmetric non-Hermitian SSH model decorated with gain and loss at its ends. Such a TWPT system works excellently at the exceptional point of the topological edge states, where the efficiency reaches the maximum, confirmed by both the analytical and experimental results. Besides, the TWPT system is less sensitive to the disorder in comparison to the conventional domino-form WPT system, owing to the topological nature of edge states. Our theoretical approach, based on the Hamiltonian theory derived from the CMT, can in principle be applied to various multi-relay WPT systems, favorably simplifying the complex mathematical calculation. Our TWPT system that acts as a 1D topological insulator with loss and gain, could serve as a versatile experimental platform to explore non-Hermitian topological physics. Finally, our work that combines the concept of WPT, topological physics, and non-Hermitian physics, paves a way towards topologically-robust and efficient long-distance WPT applications in electronics, transportation, and industry.

Note that after the submission of this work, we become aware of several independent works, which study wireless power transfer systems in a generalized SSH model theoretically [42] and experimentally [43].

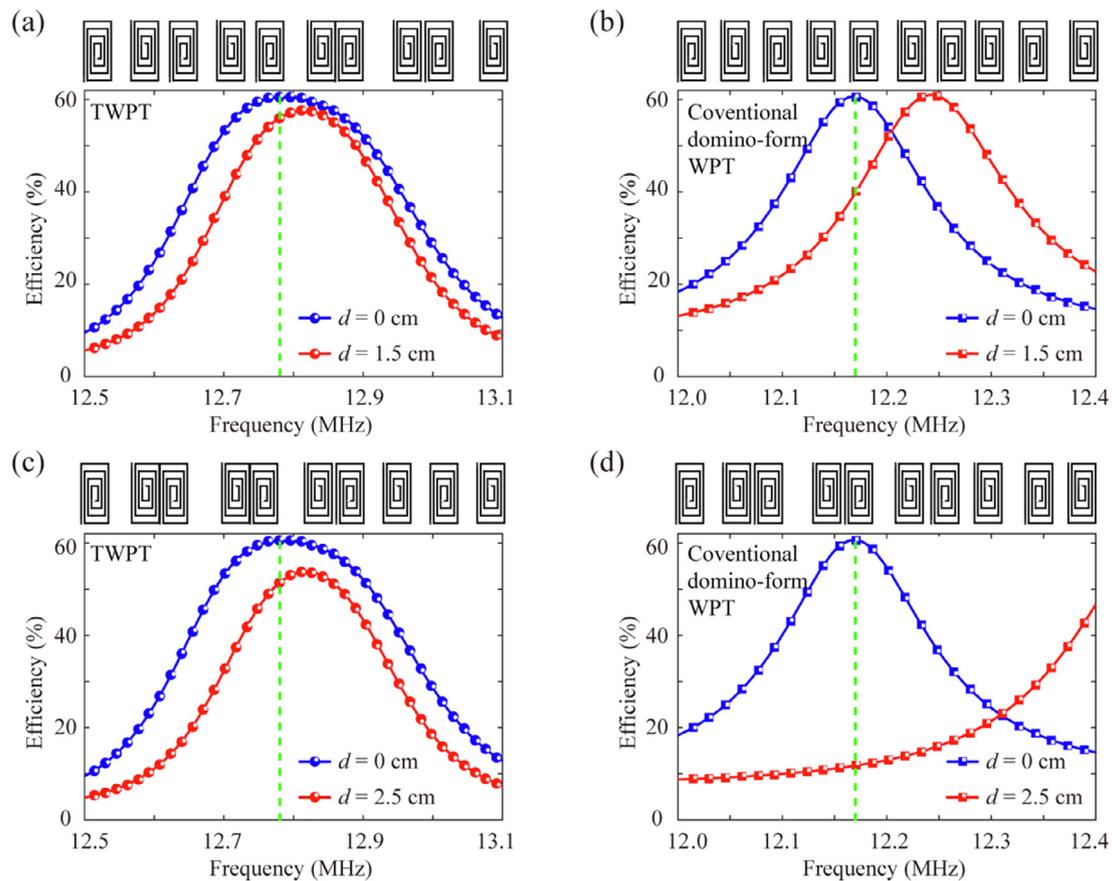

**Fig. 4.** (Color online) Experimental demonstration of the high efficiency of the TWPT system in the presence of disorder. (a)–(d) Measured transmission for the same situation with the disorder strength $d$ = 1.5 cm/ 2.5 cm in this TWPT system (a, c) and conventional domino-form WPT system (b, d). The configuration schematic is represented in the top plane. The dashed green lines represent the operating frequency $f$ = 12.78 MHz in this TWPT system (a, c) and $f$ = 12.17 MHz in conventional domino-form WPT system (b, d).

**Conflict of interest**

The authors declare that they have no conflict of interest.

**Acknowledgments**

The work at Zhejiang University was sponsored by the National Natural Science Foundation of China (61625502, 11961141010, 61975176, and U19A2054), the Top-Notch Young Talents Program of China, and the Fundamental Research Funds for the Central Universities. Work at Nanyang Technological University was sponsored by Singapore Ministry of Education under Grant Nos. MOE2018-T2-1-022 (S), MOE2015-T2-1-070, MOE2016-T3-1-006, and Tier 1 RG174/16 (S).

**Author contributions**

Yihao Yang, Li Zhang, and Hongsheng Chen conceived the original idea. Li Zhang and Yihao Yang designed the structures and the experiments. Li Zhang, Zhao Jiang, and Zhouyi Wu conducted the experiments. Li Zhang, Qiaolu Chen, Qinghui Yan, and Yihao Yang did the theoretical analysis. Li Zhang, Yihao Yang, Baile Zhang, Jiangtao Huangfu, and Hongsheng Chen wrote the manuscript and interpreted the results. Yihao Yang, Baile Zhang, Jiangtao Huangfu, and Hongsheng Chen supervised the project. All authors participated in discussions and reviewed the manuscript.

**Appendix A. Supplementary materials**

Supplementary materials to this article can be found online at https://doi.org/10.1016/j.scib.2021.01.028.

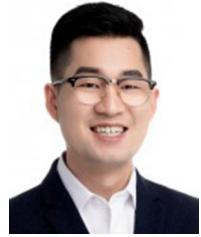

Yihao Yang received his Ph.D. degree in Electronics Science and Technology from the Department of Information Science and Electronic Engineering, Zhejiang University in 2017. He was a Research Fellow at the Center for Disruptive Photonic Technologies, Nanyang Technological University (NTU), Singapore, from 2017 to 2020. He is currently a tenure-track professor with the Electromagnetics Academy, Zhejiang University. His main research interest includes topological photonics, metamaterials, metasurfaces, and invisibility cloaks.

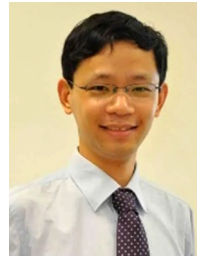

Baile Zhang received his Ph.D. degree in 2009 from Massachusetts Institute of Technology. Before he joined NTU in 2011, he was a postdoctoral associate at Singapore-MIT Alliance for Research and Technology Centre in Singapore. He is currently an associate professor at the School of Physical and Mathematical Sciences. His current research interest includes electromagnetic wave theory, invisibility cloaking, metamaterials, and acoustics.

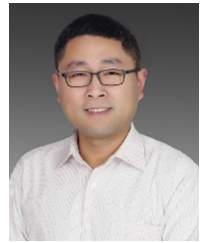

Jiangtao Huangfu received his Ph.D. degrees in Electromagnetic Field and Microwave Technology from Zhejiang University in 2004, respectively. As a visiting scholar, he worked at Massachusetts Institute of Technology in 2007 and California Institute of Technology from 2013 to 2015. From 2004, he was a lecturer and then became an associate professor in 2007 both at the Department of Information and Electronics Engineering, Zhejiang University. His recent research interest includes microwave and optical imaging, wireless sensing, radio frequency, and microwave applications.

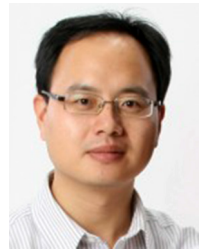

Hongsheng Chen received his Ph.D. degree in Electrical Engineering from Zhejiang University in 2005. Then, he became an Assistant Professor at Zhejiang University, an Associate Professor in 2007, and a Full Professor in 2011. He is currently a Chang Jiang Scholar Distinguished Professor with the Electromagnetics Academy, Zhejiang University. His current research interest focuses on metamaterials, invisibility cloaking, deep learning, and intelligent electromagnetic control.

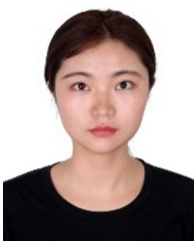

Li Zhang received her B.E. degree in Engineering from Nanjing University of Science and Technology in 2017. She is currently pursuing the Ph.D. degree at the College of Information Science and Electronic Engineering, Zhejiang University. Her current research interest includes topological insulators and non-Hermitian physics.